\documentclass[fleqn,twoside]{article}
\usepackage{espcrc2}
\usepackage{epsfig,rotating}

\def\nue{\ensuremath{\nu_{e}}}
\def\nubare{\ensuremath{\overline{\nu}_{e}}}

\def\numu{\ensuremath{\nu_{\mu}\ }}
\def\nubarmu{\ensuremath{\overline{\nu}_{\mu}}}

\def\nutau{\ensuremath{\nu_{\tau}\ }}
\newcommand{\pnuenumu}{\ensuremath{p(\nue \rightarrow \numu)\,}}
\newcommand{\nuenumu}{\ensuremath{\nue \rightarrow \numu\,}}
\newcommand{\numunutau}{\ensuremath{\numu \rightarrow \nutau\,}}

\newcommand{\nubarenubarmu}{\ensuremath{\overline{\nu}_e \rightarrow \overline{\nu}_\mu\,}}
\newcommand{\dmot}{\ensuremath{\delta m^2_{12}\,}}
\newcommand{\dmtt}{\ensuremath{\delta m^2_{23} \,}}

\newcommand{\He}{\ensuremath{^6{\mathrm{He}\,}}}
\newcommand{\Ne}{\ensuremath{^{18}{\mathrm{Ne}\,}}}

\newcommand{\thetaot}{\ensuremath{\theta_{13}}\,}
\newcommand{\thetatt}{\ensuremath{\theta_{23}}\,}
\newcommand{\numunue}{\ensuremath{\nu_\mu \rightarrow \nu_e}}

\newcommand{\sigdm}{\ensuremath{{\rm sign}(\delta m^2)}}
\newcommand{\delCP}{\ensuremath{\delta}}

\begin{document}

\title{Beta Beams}
\author{Mauro Mezzetto
\address{Istituto Nazionale Fisica Nucleare, Sezione di Padova.
Via Marzolo 8, 35100 Padova, Italy. 
{\rm E-mail:{mezzetto@pd.infn.it}}}}
\begin{abstract} 
Beta Beams could address the needs of long term  neutrino
oscillation experiments. They can produce extremely pure neutrino
beams through the decays of relativistic radioactive ions.
The baseline scenario is described, together with its physics performances.
Using a megaton water \v{C}erenkov detector installed
under  the Fr{\'e}jus, Beta Beams could improve by  a factor 200 the
present limits on $\sin^2{2 \thetaot}$
 and discover leptonic CP violating effects
if the CP phase $\delta$ would be greater than $30^\circ$ and 
\thetaot greater than $1^\circ$.
These performances can be further improved if a neutrino SuperBeam
generated by the SPL 4MW, 2.2 GeV, proton Linac would be fired to the same
detector.
Innovative ideas on higher and lower energy Beta Beams are also described.
\end{abstract}

\maketitle

\section{Introduction}

Long Baseline neutrino beams are the future facilities for neutrino
oscillations. Only in a well controlled, fully optimized environment
it will be possible to perform precision measurements of \thetatt and
\dmtt and to search for the still unknown parameters \thetaot, \delCP,
\sigdm.
These latter parameters can be explored by detecting sub leading
\numunue\  transitions, characterized by their smallness 
given the present experimental bound, from the Chooz experiment \cite{Chooz}.

The present generation of LBL neutrino experiments: K2K , Minos, Opera and Icarus
has been  designed to confirm the SuperKamiokande result on atmospheric neutrinos
through \numu\  disappearance or \numunutau transitions, with
limited sensitivity to \thetaot.

Second generation long baseline experiments, like the already 
 approved T2K \cite{T2K}, and NO$\nu$A \cite{Nova},
will extend the sensitivity on $\sin^2{2\thetaot}$ by
 more than one order of magnitude
with respect to the present Chooz limit. However they
will have very limited sensitivity to the CP phase $\delCP$ 
even if complemented by high sensitivity reactor experiments \cite{Lindner}.

A third generation of LBL neutrino experiments will be required to start a sensitive 
search of leptonic CP violation (LCPV).
These future experiments will push conventional neutrino beams to their
ultimate performances (neutrino SuperBeams), or will require new concepts in the production of
neutrino beams.

Conventional neutrino beams are generated by secondary particle decays, mainly
pions and kaons, producing a multiflavour neutrino beam (\nue, \nubarmu, \nubare, 
besides the main neutrino component, \numu).
Beam composition and fluxes are difficult to precisely predict
because of the lack of knowledge of secondary particle production
cross sections (hadroproduction). 

These limitations are overcome if the neutrino parents can be selected,
 collimated and accelerated to a given energy. The neutrino beams
from their decays would then be pure and perfectly predictable.
 This  can be tempted within the
muon or a beta decaying ion lifetimes.
The first approach brings to the Neutrino Factories \cite{Nufact},
the second to the Beta Beams.

Beta Beams  have been introduced by
P. Zucchelli in 2001 \cite{Piero}.
The idea is to generate pure, well collimated and intense
\nue\  (\nubare) beams by producing, collecting, accelerating radioactive ions
and storing them in a decay ring.
The ideal candidates are 
chosen with a lifetime around 1 s among the ions that can
 be artificially copiously produced. 
The best candidates so far are  $^{18}Ne\;$  and $^6He\;$ for \nue\ and
\nubare\  respectively.
 A baseline study for such a BetaBeam complex has been produced at
CERN \cite{Lindroos}.

In this scenario
Beta Beam neutrino energies are below 1 GeV and
the ideal detector  is a water \v{C}erenkov
detector with a mass of the order of 1 megaton,
necessary to reach the needed sensitivity.
Such a detector would have excellent physics capabilities in its own,
as fully described in reference \cite{UNO},
like ultimate sensitivities for proton decay, supernovae neutrinos, atmospheric
neutrinos etc.
A candidate site for this detector exists, located under the Frejus, at an
appropriate baseline from CERN: 130 km.

Liquid Argon could be an excellent alternative to water, provided that the
Icarus technology could be scaled to the 100 Kton scale \cite{Andre}.
\section{The BetaBeam complex}
\begin{figure}[ht]
 \epsfig{file=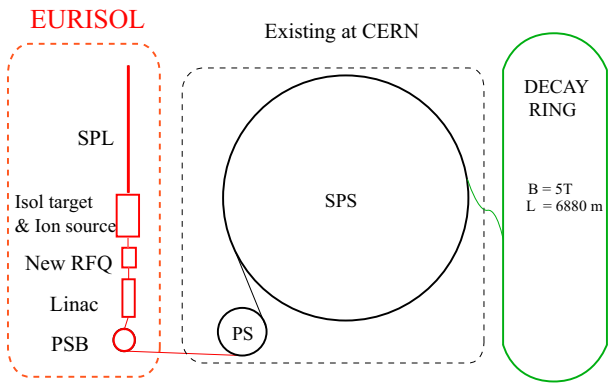,width=0.50\textwidth}  
  \vspace*{-0.9cm}
\caption{Schematic layout of the beta-beam complex. At left, the low energy part is
largely similar to the EURISOL project. The central part (PS and SPS) uses
existing facilities. At right, the decay ring has to be built.}
\label{fig:sketch}
\end{figure}

The beta-beam complex is described in \cite{Lindroos} and
 shown schematically in figure~\ref{fig:sketch}.
%
Protons are delivered by a high power Linac. Beta Beam targets need
100 $\mu$A proton beam, at energies between 1 and 2 GeV.

In case the
 Super Proton Linac (SPL) \cite{SPL} would be used,
Beta Beams could be fired to the same detector together with
a neutrino SuperBeam \cite{SPL-SB}.
 SPL 
is designed to deliver 2mA of 2.2 GeV (kinetic energy) protons, in such
a configuration Beta Beams would use 10\% of the total proton intensity,
leaving room to a very intense conventional neutrino beam.

 The targets are similar to the ones envisioned by EURISOL~\cite{Eurisol}:
 the $^6$He target consists 
either of a water cooled  tungsten core or of
a liquid lead core which works as a proton to neutron converter
surrounded by beryllium oxide \cite{Nolen}, aiming for 10$^{15}$
fissions per second. 
$^{18}$Ne can be produced by spallation reactions,
in this case protons will directly hit a magnesium 
oxide target.
The collection and ionization of the ions is performed using the ECR technique
 \cite{Sortais}. 

Ions are firstly accelerated to 
MeV/u by a Linac and to 300 MeV/u, in a single batch of 150 ns,
by  a rapid cycling synchrotron .
16 bunches (consisting of 2.5 10$^{12}$ ions each in the case of \He) are then
accumulated into the PS, and reduced to 8 bunches during their acceleration
to intermediate energies.
The SPS will finally accelerate the 8 bunches to the desired energy
 using a new 40 MHz RF system and the existing 200 MHz RF
system, before ejecting them in batches of four 10 ns bunches 
into the decay ring.
The SPS could accelerate \He ions at a maximum $\gamma$ value of
$\gamma_{\He}=150$.

The decay ring has  a total length of 6880 m
 and straight sections of 2500 m each (36\% useful length for
ion decays).
These dimensions are fixed by the  need 
to bend \He ions up to $\gamma=150$ using 5 T superconducting
magnets.
Due to the relativistic time dilatation, the ion lifetimes reach several
minutes, so that stacking the ions in the decay ring is mandatory to get enough
decays and hence high neutrino fluxes. The challenge is then to inject ions
in the decay ring and merge them with existing high density bunches.
As conventional techniques with fast elements are excluded, a new scheme 
(asymmetric merging) was specifically conceived
 \cite{mergexp}.
\subsection{Neutrino fluxes}
\label{sec:fluxes}
\Ne and \He ions can be stored together in the decay ring (of course,
in different bunches). Due to their different magnetic
 rigidities, these ions would
have relativistic $\gamma$ factors in the 5 to 3 ratio, which is quite
acceptable for the physics program. This will impose constraints on the 
lattice design for the decay ring, but no impossibility has been identified.

An ECR source coupled to an EURISOL target would 
produce $2 \cdot 10^{13}$ $^6$He ions per second.
Taking into account all decay losses along the accelerator complex, and
estimating an overall transfer efficiency of 50\%, one estimates that
$4 \cdot  10^{13}$ ions would permanently reside in the final decay ring.
That would give an antineutrino flux aimed at the Fr{\'e}jus
underground laboratory of $2.1 \cdot 10^{18}$ per standard year (10$^7$ s).

For $^{18}$Ne, the yield is expected to be $1.6\cdot 10^{12}$ ions in a 2 s
exposing time.
Due to this smaller yield, which could be certainly improved with some R\&D,
it was then proposed to use 3 EURISOL targets in sequence connected to the
same ECR source.
Again taking into account decay losses plus a 50\% efficiency, this means that
$4\cdot 10^{13}$ such ions would reside in the decay ring, 
giving rise to a neutrino flux of $0.7 \cdot 10^{18}$ per standard year.  


In the following  it was supposed that the neutrino flux from \Ne 
could be increased
by 50\% over the present conservative estimate, having
room for improvements both
in the cycle duration of PS and SPS and in
the \Ne production at the targets. A 40 \% 
improvement was put on the \He generated antineutrino fluxes. 

The reference fluxes are then
 $2.9 \cdot 10^{18}$ \He\ 
useful
decays/year and $1.1\cdot10^{18}$ \Ne\  decays/year. 
\subsection{Radiation issues}
The main losses are due to decays of He ions, and reach 1.2 W/m in the PS
and 9 W/m in the decay ring. This seems manageable, although the use of
superconducting bending magnets in the decay ring requires further studies.
Activation issues have been recently addressed \cite{Magi}, and show that
the dose rate on magnets in the arcs is limited to 2.5 mSv/h at contact after
30 days operation and 1 day cooling. Furthermore, the induced radioactivity
on ground water will have no impact on public safety.

\section{Physics reach}
Beta Beam sensitivity have  been computed assuming a water \v{C}erenkov detector
of 440 kt fiducial mass installed in the underground
Fr{\'e}jus laboratory, with  a 130 km baseline.
Most of the results of this section are taken from reference \cite{beta}.
\subsection{Signal and backgrounds} 
The neutrino beam energy is defined by
the $\gamma$ of the parent ions in the decay ring. 
 The energy optimization 
is a compromise between the advantages of the higher $\gamma$, as a
 better focusing, higher cross sections and higher signal efficiency
 and the advantages of the lower $\gamma$ 
values as the reduced background rates 
 and the better match with the
probability functions \cite{myoldbeta}.

\begin{figure}[ht]
    \begin{minipage}{\textwidth}
      \begin{minipage}{0.48\textwidth}
      {\epsfig{file=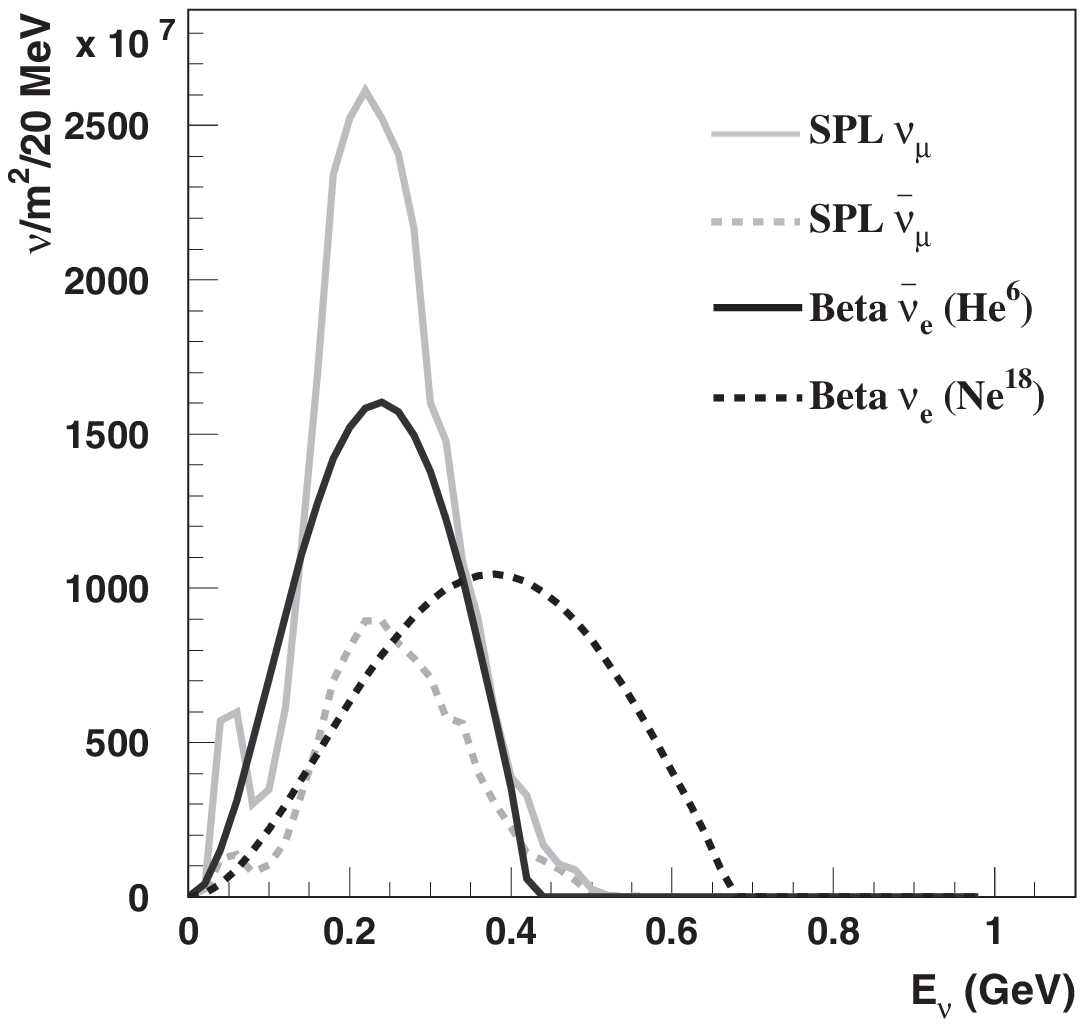,width=\textwidth}  }
      \begin{center}
      \begin{tabular}{ccc}
      \hline
           &   Fluxes       & $\langle E_\nu \rangle$  \\
           &  $\nu/m^2/yr$  & (GeV)  \\
      \hline
          \nubare ($\gamma=60$)  &   $1.97 \cdot 10^{11}$ &  0.24 \\ 
          \nue ($\gamma=100$)    & $1.88 \cdot 10^{11}$ &  0.36  \\
      \end{tabular}
      \end{center}
      \end{minipage}
    \end{minipage}
  \vspace*{-0.2cm}
    \caption{Beta Beam fluxes at the Frejus location  (130 km baseline).
     Also the SPL Super Beam \numu \  and \nubarmu \  fluxes are shown in the
    plot.}
  \vspace*{-1.2cm}
    \label{fig:fluxes}
\end{figure}

Given the decay ring  constraint (see section 2.1):
$\gamma(^6{\rm He})/\gamma(^{18}{\rm Ne})=3/5$ 
the optimal $\gamma$  values result to be $\gamma(\He)=60$ and
$\gamma(\Ne)=100$.
Fig.~\ref{fig:fluxes} shows the BetaBeam neutrino fluxes computed at a
130 Km baseline, together with the SPL Super Beam (SPL-SB) fluxes.

The mean neutrino energies of the \nubare, \nue\  beams are 0.24 GeV
 and 0.36 GeV respectively.
 They are well matched with the CERN-Frejus 130 km baseline.
On the other hand energy resolution is very poor at these energies, given
the influence of Fermi motion and other nuclear effects. 
Sensitivities are computed for a counting experiment with no
energy cuts.

A different optimization holds if each ion can be run separately at
its optimal $\gamma$ value. In this case the single fluxes can be
doubled just filling all the SPS batches with the same ion, and
in 10 years the same integrated number of useful decays/year
is obtained. Under this hypothesis the optimal choice seems to
be $\gamma_{\He}=\gamma_{\Ne}=75$.

It has been recently pointed out \cite{JJOtranto} that using
the algorithm tools discussed in section 4, Beta Beam performances
could be further improved by accelerating the ions to $\gamma_{\He}=150$
with a baseline of 300 km.

\begin{figure}[ht]
{\epsfig{file=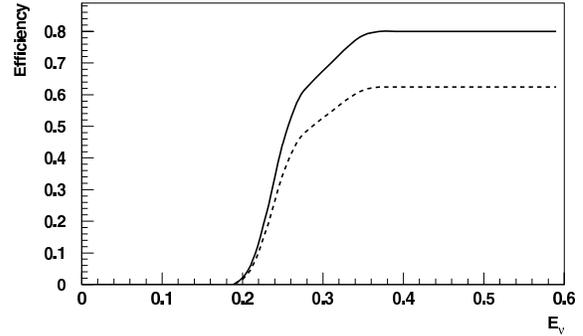,width=0.47\textwidth}  }
  \vspace*{-1.2cm}
\caption{Detection efficiency for \nubarmu (solid) and \numu\ 
(dashed) events. \numu\  efficiency is smaller given the
probability for a $\mu^-$ to be absorbed before decaying.}
\label{fig:eff}
\end{figure}

The signal in a Beta Beam looking for \nuenumu oscillations would be the
appearance of \numu\  charged-current events, mainly via quasi-elastic
interactions. These events are selected by requiring
        a single-ring event,
        the track identified as a muon using the standard 
SuperKamiokande identification algorithms (tightening the
cut on the pid likelihood value), and
        the detection of the muon decay into an electron.
Background rates and signal efficiency have been studied in a
full simulation, using the NUANCE code~\cite{casper} and reconstructing
events in a SuperKamiokande-like detector \cite{casperpriv}.
The efficiency curve for \numu\  and \nubarmu\  events
 is displayed in figure~\ref{fig:eff}.
 
The Beta Beam is intrinsically free from contamination by any different
neutrino flavor. However, backgrounds can be generated by inefficiencies
in particle identification, such as mis-identification of pions produced
in neutral current single-pion resonant interactions,
electrons (positrons) mis-identified as muons, or by external sources such as
atmospheric neutrino interactions.

The pion background has a threshold at neutrino energies of about 450 MeV,
and is highly suppressed at the Beta Beam energies. The electron background is
almost completely suppressed by the request of the detection of a delayed
Michel electron following the muon track.
 The atmospheric neutrino
background can be reduced mainly by timing the parent ion bunches.
 For a decay ring straight sections of
2.5~km and a bunch length of 10~ns, which seems feasible \cite{Lindroos}, 
this background becomes negligible \cite{Piero}.
Moreover, out-of-spill neutrino interactions can be used to
normalize it to  1\% accuracy level.

Signal and background rates for a 
4400~kt-yr exposure to  $^6$He and $^{18}$Ne beams
 are reported in Table~\ref{tab:beta:rates}.
The default values for the oscillation parameters
are \mbox{$\sin^2{2\theta_{23}}=1$}, \mbox{$\dmtt=2.5\cdot 10^{-3} {\rm eV}^2$},
 \mbox{$\sin^2{2\theta_{12}}=0.8$}, {$\dmot=7.1\cdot 10^{-5} {\rm eV}^2$},
 \sigdm=+1.
\begin{table}
\caption{\label{tab:beta:rates}
 Event rates for a 4400~kt-y exposure. The signals are
computed for \mbox{$\thetaot=1^\circ$}, 
$\delCP=90^\circ$ $\sigdm=+1$. ``$\delCP$-oscillated'' events indicates the difference between
the oscillated events computed with $\delCP=90^\circ$ and with
$\delCP=0$. ``Oscillated at the Chooz limit'' events are computed for
$\sin^2{2\theta_{13}}=0.12$, $\delCP=0$.}
\begin{tabular}{@{}lrr}
 		& $^6He$               &   $^{18}Ne$                \\
 		&       ($\gamma=60$)  &            ($\gamma=100$)  \\
\hline
CC events (no oscillation) & 19710   & 144784  \\
Oscillated (Chooz limit) & 612 & 5130   \\
Oscillated ($\delCP=90^\circ$, $\theta_{13}=3^\circ$)&  44     &  529   \\
$\delCP$ oscillated          & -9      &  5712\\
Beam background            &  0      &  0     \\
Detector backgrounds       &   1     &  397   \\
\end{tabular}
\end{table}
\subsection{Systematic errors}
Systematic errors could spoil the sensitivity of any experiment
looking for leptonic CP violations.
To this purpose a beam without any contamination 
 where the neutrino fluxes are known with great precision,
like the Beta Beam, is the ideal facility.

A major concern is the knowledge of neutrino cross-sections in an
energy range where experimental data are poor and the
cross sections are roughly proportional to $E_\nu^2$.
On the other hand Beta Beams are
the ideal place where to measure neutrino cross sections.
Being the neutrino fluxes precisely known,
a close detector of $\sim 1~$kton (fiducial) placed at a distance of 
about 1~km from the decay ring could directly measure the relevant
neutrino cross sections. 
Furthermore the $\gamma$ factor of the accelerated ions can be varied.
 In particular
an energy scan can be initiated below the background production threshold,
allowing a precise measurement of the cross sections for resonant processes.

It is estimated that a residual systematic error of 2\% will be
the final precision with which both the signal and the backgrounds
can be evaluated.

The \thetaot and $\delCP$ sensitivities are computed taking into account
        a 10\% error on the solar $\delta m^2$ and $\sin^2{2\theta}$,
  and a 5\% and 1\% error on $\delta m^2_{23}$ and $\sin^2{2\theta_{23}}$
respectively,
as expected from the T2K neutrino experiment~\cite{T2K}.
Only the diagonal contributions of these errors are considered.
\subsection{$\thetaot/\delCP$ sensitivities}
%
\thetaot can be measured either with \nue\  and \nubare\  disappearance
and with \numu, \nubarmu\  appearence.
The disappearance channels offer a cleaner extraction of \thetaot\ 
from the experimental result, but are limited by systematic
errors.
The comparison of the \nue\  and \nubare\  disappearance
measurements could also set limits to CPT violation effects.

\begin{figure}[htb]
        \centerline{\epsfig{file=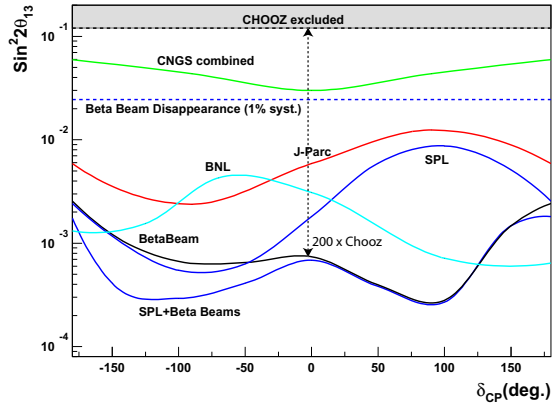,width=0.46\textwidth}  }
  \vspace*{-1.1cm}
  \caption{
   90\%CL sensitivity expressed as function of $\delCP$ for
  $\dmtt=2.5\cdot10^{-3}eV^2$.
  CNGS and J-Parc curves are taken from \cite{Migliozzi}, BNL from
  \cite{Diwan}.
   All the appearance sensitivities are computed for
 $\sigdm)=+1$ and 5 years of data taking.}
  \label{fig:th13}
\end{figure}

\thetaot and $\delCP$ are so tightly coupled in the appearance
channels 
that the sensitivity expressed for $\delCP=0$ is purely
indicative. A better understanding of the sensitivity of the BetaBeam is
expressed in the $(\thetaot,\delCP)$ plane, having fixed all the
other parameters, as shown in 
Fig.~\ref{fig:th13}).
 
\begin{figure}[ht]
\centerline{\epsfig{file=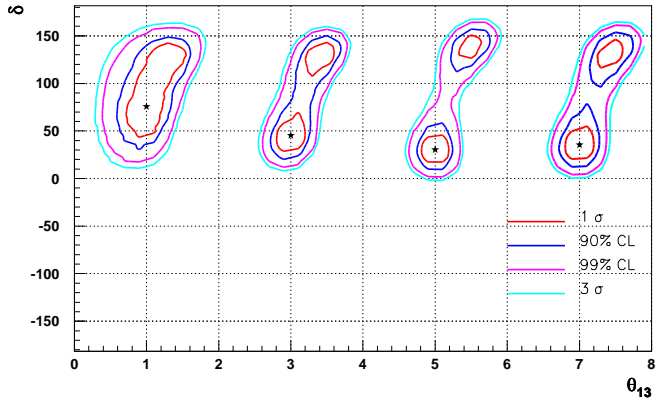,width=0.55\textwidth}  }
  \vspace*{-0.9cm}
\caption{Fits to \thetaot and \delCP\  after a 10 years 
run. Plots are shown for the (\thetaot,\delCP) values
indicated by the stars.
For the other neutrino oscillation parameters see the text.
Lines show $1 \sigma$, 90\%, 99\% and $3 \sigma$ confidence levels.  }
\label{fig:many_plots}
\end{figure}

\begin{figure}[th]
\centerline{\epsfig{file=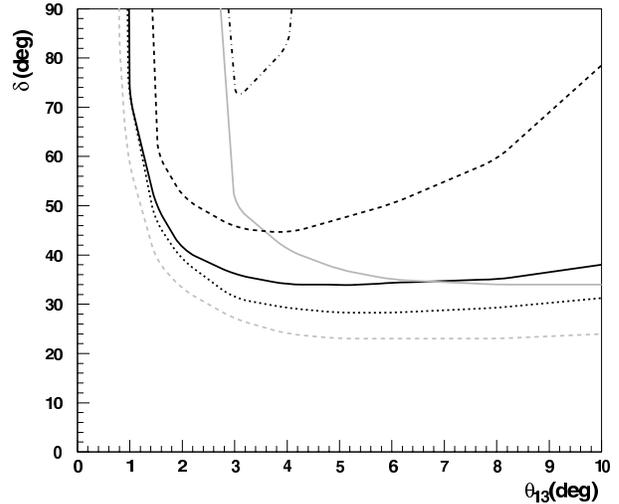,width=0.50\textwidth} }
 \vspace*{-0.9cm}
\caption{$\delCP$ discovery potential at $3 \sigma$ (see text) computed
for 10 years running time.
Black curves are  computed with 2\% systematic errors (solid),
5\% (dashed) and 10\% (dot-dashed). 
Dotted curve is the Beta Beam computed by running the two ions at
$\gamma=75$ (2\% systematic error).
 Gray solid curve is SPL-SuperBeam sensitivity,
gray dashed is the Beta Beam plus SPL Superbeam sensitivity curve,
both curves are computed with a 2\% systematic error.}
\label{fig:CP:delta}
\end{figure}

A search for leptonic CP violation  can be performed 
fitting the number of muon-like events to
the \pnuenumu and to the p(\nubarenubarmu) probabilities.
 The fit can provide the simultaneous determination
of $\theta_{13}$ and $\delCP$, see figure~\ref{fig:many_plots}.
Event rates are summarized in Table~\ref{tab:beta:rates}.
The (\thetaot,\delCP) parameter space where the experiment could
measure a $3 \sigma$ LCPV effect 
 is shown in figure~\ref{fig:CP:delta}.
Discovery potential curves are drawn for 2\%, 5\% and 10\% systematic errors.
The 5\% systematic error curve is roughly equivalent to the 2\% curve
computed with a detector having half the mass, showing 
the absolute need to keep systematic errors low
in this kind of measurements.
\subsection{Parameter correlations, degeneracies and clones}
Correlations between \thetaot and $\delCP$ are fully accounted for,
and indeed they are negligible as can be
seen in the fits to \thetaot and $\delCP$ shown in figure~\ref{fig:many_plots}.
These correlations are tiny because at the baseline of 130 km matter
effects are negligible and do not compete with genuine LCPV effects.
A full computation of degeneracies, correlations and clones for the Beta Beam
sensitivities can be found in \cite{degeneracy}, assuming that at the
time Beta Beam will start no informations about the values of \sigdm\  and
\thetatt will be available.
\subsection{Synergies between the SPL-SuperBeam and the Beta Beam}
The Beta Beam can run with the SPL as injector, but consumes at most $\sim 10\%$
of the SPL protons. The fact that the average neutrino  energies of
both the SuperBeam and the Beta Beam are below 0.5 GeV
(cfr. figure~\ref{fig:fluxes}), with the Beta
Beam tunable, offers the fascinating possibility of exposing the
same detector to $2\times 2$ beams (\numu\  and \nubarmu\  $\times$
\nue\  and \nubare) having access to
CP, T and CPT searches in the same run.

It is evident that the combination of the two beams would 
not only result result an increase of the experimental statistics,
but  would also offer    
clear advantages in the reduction of the systematic errors and 
the necessary redundancy to firmly establish any LCP
 within the reach of the experiment.

The CP violation sensitivities of the combined BetaBeam and SPL-SB 
experiments are shown in  figure~\ref{fig:CP:delta}.
\section{High energy scenarios}
It has been pointed out in \cite{HighEnergy} that allowing for higher
values of $\gamma$, that is exploiting a higher energy proton synchrotron than the SPS,
performances of the Beta Beam could be significantly enhanced.
In the paper three different scenarios are compared: the baseline scenario
described in section 2 (Setup I),
 $\gamma_{\He}=350$ with a baseline $L\sim 730~\hbox{km}$ (Setup II),
and $\gamma_{\He}=1500$ with $L\sim 3000~\hbox{km}$ (Setup III).
The same fluxes of the baseline scenario have been assumed
(the ion currents entering the decay
ring correspond to $\sim 0.5$ MW at $\gamma_{\He}=350$ and to
$\sim 2$MW at $\gamma_{\He}=1500$) and it should be noted that
the decay ring length, linearly proportional to $\gamma_{\He}$, would
be of 16 (70) km for $\gamma_{\He}=350 (1500)$ if computed with the
same parameters of the baseline scenario.

The main advantages of the most promising Setup-II scenario would be
\begin{itemize}
\setlength{\itemsep}{-4pt}
	\item
the mean neutrino energy would be $E_\nu \simeq 1.2$ GeV  and so
water \v{C}erenkov detectors are still suitable;
	\item
a factor $\sim 10$ increase in
\numu charged current event  rate (1.5 increase at constant accelerator power) with
respect to the baseline scenario;
	\item
the possibility to 
exploit energy spectrum (more powerful fits to \thetaot, $\delta$);
	\item
the possibility of measuring     sign($\Delta m^2$) (baseline $\simeq 700$ km).
\end{itemize}
On the other hand, as discussed in section 3.1, the background rate induced
by charged pion production in NC events is very high at these energies. 
However
 the longer the pion track,
the higher the probability it interacts in water missing the signature of
the decay electron. A modest cut in the visible neutrino energy can reduce
background fractions below $3\cdot 10^{-3}$ for an integrated efficiency
of $30-50\%$ \cite{HighEnergy}.

 Energy reconstruction is much more demanding at 1.2 GeV,
in particular the non quasi elastic fraction of CC events becomes  important, 
but the methods introduced  in the paper allow for an efficient
(though indirect) true 
neutrino energy estimation from the measured charged muon momentum.

\begin{figure}[th]
 \vspace*{-0.4cm}
{\epsfig{file=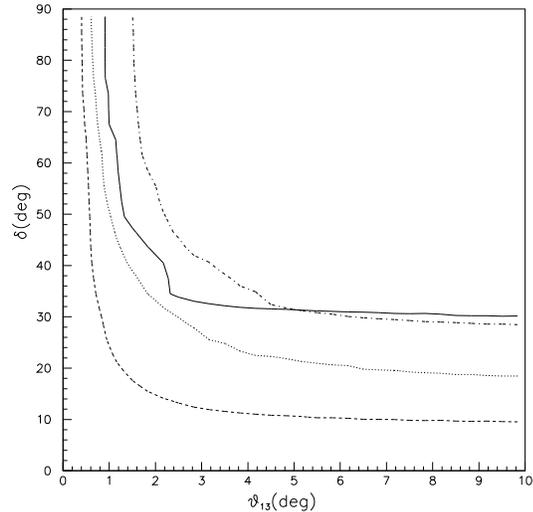,width=0.48\textwidth} }
 \vspace*{-1.2cm}
\caption{From reference \cite{HighEnergy}:
region where $\delta$ can be distinguished from $\delta=0$ or
$\delta=180^\circ$ at 99$\%$ CL for Setup-I (solid), Setup-II  with a water \v{C}erenkov detector
of 400 kton (dashed) at a baseline of 730 km and with the same detector with a factor 10 smaller mass (dashed-dotted) and Setup-III (dotted) with a 40 kton tracking calorimeter. In all cases  10 years of running time are considered.
}
\label{fig:CP_Pilar}
\end{figure}

The 99\% CL discovery potential curves in the three different scenarios,
computed including no systematic errors are shown if figure~\ref{fig:CP_Pilar}.
%
\subsection{Another  high energy scenario}
In \cite{HighEnergy2} it has been introduced the case of the
highest energy that could be accessible at LHC: $\gamma_{\He}=2488$.
It was assumed that the same fluxes of the baseline scenario
could be used and that the decay ring could be partially accommodated
inside the LHC tunnel. The neutrino fluxes at the CERN-Gran Sasso baseline
(732 km) would be so high that just counting the muons produced by
neutrino interactions in the rock, with a $15 \times 15 \;{\rm m}^2$
counter detector, the sensitivity on $\sin^2{2\thetaot}$ could be increased
by about one order of magnitude (for $\delCP=0$) with respect to the baseline scenario.
Given the neutrino energy and the baseline, the experiment would be
strongly off-peak, and no information about \delCP\  or \sigdm would
be available.
\section{Low energy scenarios}
As discussed in section 3.2, Beta Beams are the ideal place where to measure
neutrino  cross sections.
 This could allow for precision measurements 
in the range of interest of astrophysics (supernovae explosions,
nucleosynthesis etc.):
 $E_\nu=10 \; - 100 \; {\rm MeV}$, corresponding to 
$\gamma_{\He}=7\; - \; 14$.
This experimental possibility is fully discussed in \cite{Volpe}.
\section{Conclusions}
Beta Beams are a novel concept which can provide a very clean and  
powerful facility
for the search of leptonic CP violation:
 a single flavour beam and no competition
with the fake CP phenomena induced by matter effects.

The baseline scenario has been designed to be
based on available technology with some conservative extrapolations.
Its design has a very strong synergy with the EURISOL project
aiming at producing high intensity radioactive beams  
for nuclear physics studies with astrophysical applications. 

The gigantic far detector needed for these studies has excellent physics
capabilities in its own, i.e.
to study proton decay and detect supernova explosions.
A possible site exists near
CERN at the right distance.

Physics potential in neutrino oscillations would be a
$\sin^2{2\thetaot}$ sensitivity
 more than
two orders magnitude better than the present experimental limit,
having the distinctive feature to look for \thetaot also in 
\nue\  and \nubare\  disappearance channels,
 and a 3 $\sigma$ discovery potential on leptonic CP violation for
CP phase $\delta$ values greater than $30^\circ$ and for
\thetaot greater than $1^\circ$.

These performances can be further improved if a neutrino SuperBeam
generated by the SPL 4MW, 2.2 GeV, proton Linac would be fired to the same
detector. In this configuration the two beams could address leptonic
 CP and T violation, and could also explore
CPT violation in neutrino oscillations.

Several new ideas about the baseline option optimization and extensions
to higher and lower energies have been recently published,
showing the great and partially still unexplored potential of the 
Beta Beams.

%
%
%
%
%

\end{document}